\documentclass[]{elsart}
\usepackage{natbib}
\usepackage{graphicx}
\usepackage{amssymb}
\usepackage{amsmath}
\usepackage{times}
\usepackage{ulem}

\begin{document}

\begin{frontmatter}
\title{Strategy abundance in $\boldsymbol{ 2 \times 2}$ games\\ for arbitrary mutation rates}

\author[PED]{Tibor Antal},
\author[PED]{Martin A. Nowak},
\author[MP]{Arne Traulsen}
\address[PED]{Program for Evolutionary Dynamics, Harvard University, Cambridge MA 02138, USA}
\address[MP]{Max-Planck-Institute for Evolutionary Biology, 24306 Pl{\"o}n, Germany}

\begin{abstract}
We study evolutionary game dynamics in a well-mixed populations of finite size, $N$. A well-mixed population means that any two individuals are equally likely to interact. In particular we consider the average abundances of two strategies, $A$ and $B$, under mutation and selection. The game dynamical interaction between the two strategies is given by the $2 \times 2$ payoff matrix $\bigl(\begin{smallmatrix} a & b\\ c & d \end{smallmatrix}\bigl)$.
It has previously been shown that $A$ is more abundant than $B$, if $a(N-2)+bN>cN+d(N-2)$.
This result has been derived for particular stochastic processes that operate either in the limit of asymptotically small mutation rates or in the limit of weak selection. Here we show that this result holds in fact for a wide class of stochastic birth-death processes for arbitrary mutation rate and for any intensity of selection.
\end{abstract}

\end{frontmatter}

\section{Introduction}

Evolutionary dynamics describe how successful strategies spread in a population through genetic reproduction or cultural imitation. In mutation-selection processes where the fitness of each individual is constant, the competition between two types $A$ and $B$ is straightforward.
If both fitness values are identical, then we have neutral evolution \citep{kimura:1968ab} and the average abundances of $A$ and $B$ are the same. (Throughout the paper we assume symmetric mutation rates: the mutation probability from $A$ to $B$ is the same as from $B$ to $A$.) If $A$ is fitter than $B$, then $A$ is more abundant than $B$ in the mutation selection equilibrium. Thus, for constant selection the comparison of the abundances of the two types is trivial. 

For frequency dependent selection, where fitness depends on the types and abundances of others, the situation is more complex. One way to model these systems are evolutionary games, where the reproductive success of individuals depends on their payoff derived from interaction with others  \citep{maynard-smith:1973to,maynard-smith:1982to,weibull:1995hp,samuelson:1997sa,hofbauer:1998mm,fudenberg:1998bv,hofbauer:2003mm,nowak:2004aa,szabo:2007aa,sandholm:2007bo}. 
In the simplest case, interactions are described by a $2 \times 2$ payoff matrix, 
\begin{equation}
\label{payoffmatrix}
\bordermatrix{
  & A & B \cr
A & a & b \cr
B & c & d \cr}.
\end{equation}
Here, an $A$ individual obtains $a$ from other $A$ individuals, but $b$ from $B$ individuals. 
Similarly, $B$ obtains $c$ from $A$, and $d$ from other $B$ individuals.

For $a>c$ and $b>d$, strategy $A$ dominates strategy $B$: Regardless of the composition of the population, strategy $A$ has the higher payoff. Thus, evolutionary dynamics will always lead to a population with more $A$ individuals than $B$ individuals. Equivalently, for $a<c$ and $b<d$, strategy $B$ dominates. In this case, the average abundance of $B$ will be higher than the average abundance of $A$ at the mutation-selection equilibrium.

For $a>c$ and $b<d$, both strategies are best replies to themselves. This is a `coordination game'. In a population of mostly $A$ individual, rare $B$ mutants have a lower payoff. In a population of mostly $B$ individuals, rare $A$ mutants have a lower payoff. Therefore, in a mutation selection process it is not a priori clear whether strategy $A$ or strategy $B$ will be more abundant at the equilibrium distribution. Two concepts are important: (i) pareto efficiency and (ii) risk-dominance.  
Strategy $A$ is Pareto efficient if $a>d$. In this case, an all-$A$ population has a higher average payoff than an all-$B$ population. Strategy $A$ is risk-dominant if $a+b>c+d$ \citep{harsanyi:1988mm}. In this case, strategy $A$ has a larger basin of attraction than strategy $B$. In a game between two players, if I do not know what the other person will do, it is less risky for me to choose the risk dominant strategy, but it would be more rewarding for both of us to choose the Pareto efficient strategy. A coordination game is especially interesting, if the risk-dominant strategy is not Pareto efficient.

\citep{kandori:1993aa} have shown that $A$ is chosen over $B$ if $a(N-2)+bN>cN+d(N-2)$. 
This means a large population selects the risk dominant equilibrium in the long run. They analyze a process which is stochastic in the generation of mutants, but deterministic in following the gradient of selection. Their calculation assumes asymptotically small mutation rates, but holds for a wide range of evolutionary processes, in contrast to previous approaches that make specific assumptions on the source of noise \citep{foster:1990bv,fudenberg:1992bv}.
However, since the dynamics follows the gradient of selection, the processes can only leave a Nash equilibrium if a sufficient number of mutations occurs simultaneously. In large populations, it is very unlikely that this happens, and the time until the population moves from one equilibrium to another becomes exponentially large in $N$.

If the dynamics is stochastic, a single mutation can be sufficient to leave a Nash equilibrium. 
\cite{nowak:2004pw} have studied the competition of two strategies $A$ and $B$ in a frequency dependent Moran process (and similar processes) which model fully stochastic evolutionary dynamics. This means that the selection steps are stochastic. They have shown that the fixation probability of $A$ is greater than that of $B$ if and only if  $a(N-2)+bN>cN+d(N-2)$. This calculation assumes weak selection in a process with selection only.
The result is also valid for stronger intensities of selection, if we use the pairwise comparison process \citep{szabo:1998wv,traulsen:2007cc}
or a slightly modified version of the Moran process
for evolutionary updating \citep{traulsen:2008aa}.

Here, we extend these results to both arbitrary mutation rates and arbitrary intensities of selection. We show that $A$ is more abundant than $B$ in the mutation-selection distribution for a wide range of stochastic processes for any mutation rate and any intensity of selection if and only if 
$a(N-2)+bN>cN+d(N-2)$.
Higher mutation rates seem to be very relevant for social evolution, because humans (and higher animals) tend to try new strategies fairly often. We therefore expect that evolutionary dynamics in the cultural learning and imitation context operate under fairly high mutation (=`exploration') rates. We also show that the condition holds for any payoff matrix, not only for coordination games.

\section{Pairwise comparison process}

The payoffs of $A$ and $B$ in a population of $j$ individuals 
of type $A$ and $N-j$ individuals of type $B$ are 
\begin{equation}
\label{payoff}
\begin{split}
\pi_A(j) &= \frac{j-1}{N-1} a + \frac{N-j}{N-1} b 
\\ 
\pi_B(j) &= \frac{j}{N-1} c + \frac{N-j-1}{N-1} d .
\end{split}
\end{equation}
Each individual interacts with $N-1$ other individuals.  
The payoff difference $\Delta  \pi(j)$ is a linear function of $j$
\begin{equation}
\label{deltapi}
\Delta  \pi(j)=\pi_A(j)-\pi_B(j) =  \frac{a-b-c+d}{N-1} j + \frac{-a+bN-dN+d}{N-1}.
\end{equation}

First, we need to specify how strategies spread in the population, depending on
their payoffs. 
We adopt the following process:
a random (focal) individual $i$ is selected.
It compares its payoff $\pi_i$ to the payoff 
$\pi_j$ of a randomly chosen role model, individual $j$, 
and takes over the strategy of that individual with probability  
$\left[1+e^{ \beta (\pi_i - \pi_j)} \right]^{-1}$
\citep{blume:1993jf,szabo:1998wv,traulsen:2007cc,sandholm:2007bo}. 
This process occurs with probability $1-\mu$. 
With probability $\mu$, a mutation occurs and the focal individual produces 
an offspring with random strategy, $A$ or $B$.
The quantity $\beta$ determines the intensity of selection ($0 \leq \beta < \infty$). 
Strong selection (large $\beta$) means that 
the probability to adopt a better strategy 
approaches one, and the probability to adopt
an inferior strategy vanishes. For weak selection (small $\beta$), these two probabilities are close to $1/2$. The transition probabilities to increase or decrease the number of $A$ players  by one, are given by
\begin{equation}
\label{transprob}
\begin{split}
T_{j}^+ &= \frac{j}{N} \frac{N-j}{N}
\frac{1-\mu}{1+e^{- \beta \Delta  \pi(j)}}
+ \frac{N-j}{N} \frac{\mu}{2}
\\
T_{j}^- &= \frac{j}{N} \frac{N-j}{N} \frac{1-\mu}{1+e^{+ \beta \Delta  \pi(j)}}
+ \frac{j}{N} \frac{\mu}{2} ~.
\end{split}
\end{equation}
With these transition probabilities, we can determine whether $A$ is more abundant than $B$ or vice versa. 
Before discussing the problem for general mutation rates 
we first consider the special case of low mutation rates, because of its importance in previous papers
\citep{kandori:1993aa,nowak:2004pw,taylor:2004wv,imhof:2006ee}.
We will also present a simple approximation for very high mutation rates, which illustrates the dynamics in this extreme limit.

\subsection{Low mutation rates}
\label{low}

First, we consider low mutation rates, $\mu \ll N^{-2}$. In this limit, a mutation reaches extinction or 
fixation before a second mutation arises. Thus, the important quantity is the probability that a single mutant takes over the population in a process without mutations \citep{nowak:2004pw,taylor:2004wv,imhof:2006ee,antal:2006aa}. The ratio of the fixation probability $\rho_A$ of a single $A$ player and that of a single $B$ player can be written as \citep{karlin:1975xg,nowak:2006bo}
\begin{equation}
\frac{\rho_{B}}{\rho_{A}} = {\prod_{j=1}^{N-1} \frac{T^-_j}{T^+_j}} ~.
\end{equation}
Without mutations, $\mu=0$, the ratio of the transition probabilities \eqref{transprob}, simplifies to 
${T^-_j}/{T^+_j} =e^{- \beta \Delta  \pi(j)}$
\citep{traulsen:2006bb,traulsen:2007cc}. 
Thus, ${\rho_{B}}/{\rho_{A}}$ reduces to 
\begin{equation}
\frac{\rho_{B}}{\rho_{A}} 
= \exp \left[-\beta \sum_{j=1}^{N-1}  \Delta  \pi(j)  \right]
= \exp \left[-\beta \left(\frac{a+b-c-d}{2}N -a+d \right) \right].
\end{equation}
Therefore, $\rho_{A} = \rho_{B}$ is equivalent to 
\begin{equation}
(a+b-c-d)N =2a-2d ~.
\label{condequal}
\end{equation}
From the payoff matrix \eqref{payoff} it is clear, that strategy $A$ is favored whenever $a$ or $b$ increases, or $c$ or $d$ decreases. 
Therefore, $A$ is more abundant than $B$ if
\begin{equation}
a(N-2)+b N > c N+d(N-2) ~.
\label{cond1}
\end{equation}
This condition has been derived previously in the low mutation limit for different evolutionary processes \citep{kandori:1993aa,nowak:2006bo,traulsen:2008aa}. 
The strategy with the higher fixation probability is the more abundant one, because more of its mutants reach fixation. Thus for low mutation rates, $A$ has a higher abundance than $B$ whenever condition \eqref{cond1} is fulfilled.

For weak selection, a wide range of processes fulfills the $1/3$-rule, which states that the fixation probability of a single $A$ mutant in a coordination game is larger than $1/N$ if 
$a (N-2) + b (2N-1) >c (N+1)+ d (2N-4)$ \citep{nowak:2004pw,imhof:2006aa,lessard:2007aa,Bomze:2008lr}. 
The fixation probability for large $N$ can thus be written as 
\begin{equation}
\label{Afix}
\rho_A = \frac{1}{N} + \omega \left(  \frac{1}{3} - \frac{d-b}{a-b-c+d} \right),
\end{equation}
where $\omega $ is a small positive number that may depend on the payoffs and on $N$. Writing down the analogous equation for $\rho_B$ and subtracting yields
\begin{equation}
\rho_A  - \rho_B 
=
\omega \left(  \frac{a-c}{d-c-b+a} - \frac{d-b}{a-b-c+d} \right) 
\propto a+b  - c- d ~.
\end{equation}
Thus, $\rho_A > \rho_B$ is equivalent to $a+b > c+d$, which is identical to our condition \eqref{cond1} for large $N$. 
\cite{lessard:2007aa} have shown that any process within the domain
of Kingman's coalescent \citep{kingman:1982aa} fulfills the 1/3-rule
(for large $N$ and weak selection and mutation). 
%In a nutshell, under Kingman's coalescence the number of offspring may not differ too much among individuals.
If an evolutionary process fulfills the 1/3-rule, $\rho_A$ and $\rho_B$ can be written in the form of
\eqref{Afix}. Hence we find again that $\rho_A > \rho_B$
is equivalent to $a+b>c+d$. Thus, we conclude that
our result is valid for any process within the domain of Kingman's coalescent
for large $N$ and weak selection and mutation. 

\subsection{High mutation rates}
\label{high}

For high mutation rates $\mu \to 1$, mutations dominate the process, and  
drive the system towards equal abundance of $A$ and $B$, that is close to $j=N/2$. 
Hence the relevant transition probabilities are $T^+_{N/2}$ and $T^-_{N/2}$,
and it is plausible to assume that $T^+_{N/2} > T^-_{N/2}$ implies that $A$ is more abundant than $B$ in the stationary state. 
From \eqref{transprob} we obtain
\begin{equation}
T_{N/2}^+ - T_{N/2}^-  = \frac{1-\mu}{4} \tanh \left( \frac{ \beta}{2} \Delta  \pi({N/2}) \right),
\end{equation}
where the payoff difference is given by
\begin{equation}
\Delta  \pi({N/2}) 
= \frac{(a+b-c-d)N-2(a-d)}{2(N-1)}.
\end{equation}
Since $\Delta  \pi({N/2})>0$ implies $T_{N/2}^+ - T_{N/2}^- >0$, it also implies a more abundant $A$, for arbitrary intensity of selection $\beta$.
Thus, we have again the same condition $a(N-2)+bN>cN+d(N-2)$ for the dominance of $A$.

\subsection{Arbitrary mutation rates}
\label{arbitrary}

So far we have shown that both for low and high mutation rates the same condition %\eqref{cond1} 
determines whether $A$ or $B$ is more abundant, regardless of the intensity of selection. Now, we turn to general mutation rates. 
Whenever equation \eqref{condequal} holds, 
the payoff difference \eqref{deltapi} has the following symmetry property
\begin{equation}
\Delta  \pi(j) = -\Delta  \pi({N-j}) ~.
\end{equation}
This antisymmetry in the payoff differences then implies 
the following symmetry in our transition probabilities \eqref{transprob},
\begin{equation}
\label{transsym}
 T^+_j = T^-_{N-j} ~.
\end{equation}
For symmetric transition probabilities, however, the stationary probabilities $p_j$ of having $j$ number of A players, are also symmetric,  $p_j=p_{N-j}$.
This can be shown by first writing the stationary probability distribution explicitly \citep{kampen:1997xg,claussen:2005eh}
\begin{equation}
\label{stacdef}
p_j = p_0 \prod_{i=0}^{j-1} \frac{T^+_i}{T^-_{i+1}}
\end{equation}
where $p_0$ follows from the normalization $\sum_{j=0}^N p_j =1$.
For symmetric transition probabilities \eqref{transsym} and for $j<N-j$ we can write
\begin{equation}
 \frac{p_{N-j}}{p_j} = \prod_{i=j}^{N-j-1} \frac{T^+_i}{T^-_{i+1}}
  = \prod_{i=j}^{N-j-1} \frac{T^+_i}{T^+_{N-i-1}} = 1.
\end{equation}
Hence the distribution is symmetric $p_j=p_{N-j}$.

For such a symmetric distribution, %\eqref{distsym}, 
the number of $A$ and $B$ players are of course the same, since
\begin{equation}
 \langle j \rangle = \sum_{j=0}^N jp_j = \sum_{j=0}^N jp_{N-j}  = \sum_{j=0}^N (N-j)p_j
 = \langle N-j \rangle ~.
\end{equation}
Hence we have demonstrated that when \eqref{condequal} is fulfilled, then the average abundance of $A$ players and $B$ players are the same. 

On the other hand, when condition \eqref{cond1} is fulfilled, then the average abundance of $A$ players is higher than the average abundance of $B$ players. This is quite obvious by looking at the payoff matrix \eqref{payoffmatrix}: increasing $a$ or $b$ favors $A$, while increasing $c$ or $d$ favors $B$. 
More formally, we could say that from \eqref{payoff},  \eqref{deltapi} and \eqref{transprob} 
it is clear that $T^+_j$ and $-T^-_j$ are monotone increasing functions of $a$ and $b$, and decreasing functions of $c$ and $d$. This implies condition \eqref{cond1}.
Interestingly, other features of the game do not matter, e.g. whether a single $A$ player has a larger disadvantage in a $B$ population than a single $B$ player in an $A$ population. 

Condition \eqref{cond1} can also be written in the form 
\begin{equation}
\label{finiteN}
a \left(1-\frac{2}{N} \right)+b>c+d \left(1-\frac{2}{N} \right),
\end{equation}
which highlights the fact that there is a $1/N$ correction compared to simple risk dominance ($a+b>c+d$), which is the $N\to\infty$ limit. %This argument has already been made by \citep{kandori:1993aa}.
This implies that a strategy can be more abundant than the other strategy for large $N$,  but less abundant for small $N$. For example for the payoff matrix $\bigl(\begin{smallmatrix} 10 & 3\\ 11 & 1 \end{smallmatrix}\bigl)$, strategy $A$ is more abundant in large  populations $(N\ge 19)$, but $B$ is more abundant in small  populations $(N\le 17)$. They are equally abundant for $N=18$. %Conversely, for the payoff matrix $\bigl(\begin{smallmatrix} 1 & 11\\ 3 & 10 \end{smallmatrix}\bigl)$, strategy $A$ is only abundant for small $(N\le 17)$ population sizes. 
In general, there is a threshold population size $N^*=2(a-d)/(a-d+b-c)$. If $N^*>2$, then $A$ is more abundant than $B$ either for $N>N^*$ or for $N<N^*$.

\subsection{Including self-interactions}

So far, we have adopted the usual convention that individuals cannot interact with themselves. If instead we allow individuals to derive a payoff from self-interaction, we obtain
\begin{equation}
\label{payoffSelf}
\begin{split}
f_A(j) &= \frac{j}{N} a + \frac{N-j}{N} b 
\\ 
f_B(j) &= \frac{j}{N} c + \frac{N-j}{N} d.
\end{split}
\end{equation} 
Hence, the payoff difference is $\Delta  \pi(j)=  (a-b-c+d) \, j/N + b-d$. 
Now for all mutation rates and all intensities of selection the condition for the average abundance of $A$ to exceed the average abundance of $B$ is simple risk dominance:
\begin{equation}
\label{selfcond}
a+b>c+d ~.
\end{equation}
We conclude that the finite $N$ correction to risk dominance in \eqref{finiteN} results from the exclusion of self-interactions.

\section{Frequency dependent Moran process}

Although condition \eqref{cond1} is valid for a large class of evolutionary processes (see Sec.~\ref{general}), it does not always hold if we depart from weak selection.  As an example we discuss the frequency dependent Moran process \citep{nowak:2004pw,taylor:2004wv,antal:2006aa}.  As we shall see, for this model our condition \eqref{cond1} is only valid in the limit of weak selection.

Let the fitness be a convex combination of a background fitness (which we set to $1$) and the payoff,
$f_A(j)= 1-w+w \pi_A(j)$ and $f_B(j)= 1-w+w \pi_B(j)$. Here, $w$ is the intensity of selection ($0 \leq w \leq 1$ for payoff matrices with positive entries).
An individual is selected for reproduction at random, but proportional to fitness.
The selected individual produces an offspring, which replaces a randomly chosen individual. %, and 
Mutation can occur during reproduction. This leads to the transition probabilities
\begin{equation}
\label{Moran}
\begin{split}
T_{j}^+ &= \frac{j \, f_A(j) }{F(j)} \frac{N-j }{N} (1-\mu)
+ \frac{(N-j) f_B(j) }{F(j)} \frac{N-j }{N} \mu \\
T_{j}^- &= \frac{(N-j) f_B(j) }{F(j)} \frac{j }{N}(1-\mu)
+ \frac{j \, f_A(j) }{F(j)}\frac{j}{N}\mu ~.
\end{split}
\end{equation}
Here $F(j) = j f_A(j) +(N-j) f_B(j)$ is the total fitness of the whole population.
These transition probabilities depend on the fitness values only 
through the ratio $f_A(j)/f_B(j)$. Yet, there is no 
simple condition for the equilibrium abundance of $A$ players.
In the weak selection limit, $w \ll 1$, however,
only the payoff differences enter into the transition probabilities \eqref{Moran},
\begin{equation}
 \frac{f_A(j)}{f_B(j)} = 1+ w\Delta  \pi(j) + \mathcal{O}(w^2) ~.
\end{equation}
Thus, for weak selection, condition \eqref{cond1} again ensures the higher abundance of $A$. 

We can also consider a variant of the frequency dependent Moran process where fitness is an exponential function of payoff,
$f_A(j)= \exp[+w \pi_A(j)]$ and $f_B(j)= \exp[+w \pi_B(j)]$
 \citep{traulsen:2008aa}. In this case, the transition probabilities again only depend on the payoff differences, because 
 $f_A(j)/f_B(j)=\exp(w \Delta  \pi(j))$. Hence, the condition for abundance is again \eqref{cond1} for any intensity of selection and any mutation rate.

\section{General birth death processes}
\label{general}

Here, we show that our finding holds for a wide range of birth-death processes. We need two requirements to be fulfilled. The first is that the payoffs received from other players should be additive. In this case the payoffs $\pi_A(j)$, $\pi_B(j)$, and also the payoff difference $\Delta  \pi(j)$ are linear functions of $j$. Consequently,  
\begin{equation}
 \Delta  \pi(j) + \Delta  \pi({N-j}) = \lambda(a,b,c,d,N)
\end{equation}
is independent of $j$. 
Hence, when $\lambda(a,b,c,d,N)=0$ holds, we have an antisymmetric payoff function
\begin{equation}
\label{anti}
 \Delta  \pi(j) = -\Delta  \pi({N-j}).
\end{equation}

The second requirement is that the difference between the two types of players should manifest itself in the transition probabilities $T_j^+$ and $T_j^-$, only through the the payoff difference $\Delta  \pi(j)$. 
Hence, we can write $T^-_j = T^-_j[\Delta  \pi(j), \mu]$, and the probability of decreasing the number of mutants is
\begin{equation}
\label{firststep}
 T^-_j =T_j^-[\Delta \pi(j), \mu]= T^+_{N-j}[ -\Delta  \pi(j), \mu].
\end{equation}
That is, the only difference between the two types is the change of sign of the payoff difference $\Delta  \pi(j)$. Otherwise, the transition probabilities are the same, but with the number of the opposite type of individuals $N-j$.
Now, simply setting the index to $N-j$ in \eqref{firststep}, we have
\begin{equation}
 T^-_{N-j} = T^+_j[ -\Delta  \pi({N-j}), \mu] =  T^+_j[\Delta  \pi({j}), \mu] =  T^+_j.
\end{equation}
For the second equality we have used \eqref{anti}. These symmetric transition probabilities then imply an equal abundance of the two types of players in any birth-death process, as it has been shown in Section~\ref{arbitrary}. The condition for a higher abundance of $A$ compared to $B$ is then $\lambda(a,b,c,d,N)>0$. 
This general condition takes the form of \eqref{cond1} or \eqref{selfcond} in our examples when self-interaction are excluded or included, respectively. 
Note that similar arguments appear in \citep{claussen:2007}.

\section{Discussion}
\label{disco}

In this paper, we have shown that $a(N-2)+b N > c N+d(N-2)$ is the crucial condition for $A$ to be favored over $B$ in wide range of evolutionary processes for any mutation rate, any intensity of selection and any finite population size. By $A$ being favored over $B$ we mean that $A$ is more abundant than $B$ in the mutation-selection equilibrium of the stochastic process.

The relative abundance in the stationary state is a natural way of comparing two strategies. In the low mutation limit, when the population is typically homogeneous, a strategy being more abundant is equivalent of having a larger fixation probability. Hence, our study of abundance can be viewed as a generalization of that concept to arbitrary mutation rates. This is of special interest for cultural dynamics, where mutation rates are not necessarily small.

In particular, we have generalized the famous result of  \cite{kandori:1993aa} to any mutation rate. 
We have generalized the result of \cite{nowak:2004pw} to any intensity of selection.
Our results are valid for any mutation rate and any intensity of selection.

It turns out that for our result to hold, a birth-death process has to fulfill two requirements: (i) additive payoffs, and (ii) that that the evolutionary dynamics depend only on the payoff differences (the players are identical otherwise).
These requirements hold, for example, for the pairwise comparison process described by Traulsen et al.\ (2006) and for a frequency dependent Moran process with exponential fitness function (Traulsen et al, 2008). For the standard frequency dependent Moran process with linear fitness function these requirements only hold in the limit of weak selection.

Finally, we note that in coordination games with strong selection both the fixation probability of $A$ and the fixation probability of $B$ are very small. For small mutation rates, the system will stay in one equilibrium for a very long time. Therefore, it can take a very long time for the system to obtain a representative sample of the stationary mutation-selection distribution.

\section*{Acknowledgments} We are grateful for financial support from NIH Grant R01GM078986, The John Templeton Foundation, Jeffrey Epstein (T.A. and M.A.N.) and the Emmy-Noether program of the DFG (A.T.).

%\bibliographystyle{elsart-harv}
%\bibliography{/Users/arne/traulsen}

\end{document}